\documentstyle[preprint,revtex,12pt]{aps}

\begin{document}
\begin{title}
\begin  {center}
Number of solutions of the TAP equations for\break
p--spin interaction spin glasses
\end{center}
\end{title}

\author{H.\ Rieger\cite{hrieger}}
\begin{instit}
Institut for Theoretical Physics\\
University of California\\
Santa Barbara, CA~93106.
\end{instit}
\begin{abstract}
The number $\langle N_s\rangle$ of solutions of the equations of
Thouless, Anderson and Palmer for p--spin interaction spin
glass models is calculated. Below a critical temperature $T_c$ this
number becomes exponentially large, as it is in the SK--model ($p=2$).
But in contrast to
this, for any $p>2$ the factor $\alpha(T)=N^{-1}\ln\langle N_s\rangle$
jumps discontinuously at $T_c(p)$, which is consistent with the
discontinuity occuring within the mean--field theory for these models.
For zero temperature the results obtained by Gross and M\'ezard are
reproduced, and for $p\rightarrow\infty$ one gets the result for
the random energy model.
\end{abstract}

\begin{center}
Accepted for publication in {\em Phys.~Rev.~B}
\end{center}

\vskip 1.0 cm
\narrowtext
\newpage

\section{Introduction}\label{sec1}

The free energy landscape of a spin glass \cite{by}
has an extremely complicated structure, which manifests itself in the
presence of many valleys with high barriers between them. Below a
critical temperature $T_c$ the number of valleys becomes exponentially
large and the barriers diverge with system size (at least in the
mean--field limit). The local mean--field equations for the
site--magnetizations, originally  written down by Thouless, Anderson
and Palmer (TAP) \cite{tap} in an effort to solve the
Sherrington--Kirkpatrick (SK) model \cite{sk} without using
replicas, give rise to an exponetially large number $N_s$ of solutions
below $T_c$. In the case of the SK model it was shown \cite{bray}
that $N_s\propto\exp[\alpha(T)\,N]$, where $\alpha(T)$ increases
very slowly ($\alpha(T)\propto(T-T_c)^6$) below $T_c$. To recover
the Parisi--solution \cite{parisi} of the SK model one has to
attribute Boltzman weights to each of these solutions before taking
the average over the bond--distribution \cite{domyoung}, since
only those with the lowest free energies contribute significabtly
to the thermodynamics.

The spin glass transition occuring in the mean--field theory
of p--spin interaction spin glasses
\cite{derrida,gross,gardner,sommers} is somewhat different
for $p>2$ (for a discussion of finite--dimensional
realizations of these models see \cite{hrtrk,hr}). The equilibrium
phase transition at a temperature $T_K$ manifests itself by a
discontinuous jump in the EA order parameter (although the transition
itself is of second order). But already at a higher temperature
$T_g>T_K$ a discontinuous dynamical freezing transition takes place,
where spin--autocorrelations do not decay any more on finite time
scales \cite{trk}. Within the TAP approach this means that at $T_g$
an exponentially large number of solutions has to appear, which are
uncorrelated and have a higher free energy than the paramagnetic
state --- similar to what happens in p--state Potts glasses with
$p>4$ \cite{trkwol}.

In this paper we generalize the above mentioned calculation of
Bray and Moore \cite{bray} to the case of p--spin interaction spin
glasses. We perform a ``white average'' \cite{dometal}, so one
cannot discuss the thermodynamics of these models on the basis of
our calculation. Nevertheless it will be interesting to observe
that the temperature dependence of $N_s(T)$ is essentially different
for $p>2$ from that of the SK model --- in a way that is consistent
with a discontinuous transition. The organization of the paper is as follows:
In section \ref{sec2} we formulate the problem and derive the
self--consistency equations, which are solved in section \ref{sec3},
where also the results are presented. Section \ref{sec4} contains
their discussion and two lengthy calculations are deferred to the
appendices.

\section{Number of TAP--solutions}\label{sec2}

The Hamiltonian for $p$--spin interaction spin glasses within the
mean field approximation reads
\begin{equation}
{\cal H}=-\sum_{i_1<\cdots<i_p} J_{i_1\cdots i_p}
\sigma_{i_1}\cdots\sigma_{i_p}-H\sum_i\sigma_i\;,
\end{equation}
$\sigma_i=\pm1$, $i=1,\ldots,N$ and $H$ is an external field.
The $p$--spin couplings are quenched random variable distributed
according to a Gaussian
\begin{equation}
{\bf P}(J_{i_1\cdots i_p}) = {1\over{\sqrt{\pi} {\tilde J}}}
\exp\biggl(-{J_{i_1\cdots i_p}^2\over{{\tilde J}^2}}\biggr)
\quad,\qquad{\tilde J}^2={J^2p!\over{N^{p-1}}}\;.\label{e2}
\end{equation}
Each spin feels a local field of strength
\begin{equation}
h_i={1\over{(p-1)!}}\sum_{j_2,\ldots,j_p} J_{ij_2\cdots j_p}
\sigma_{j_2}\cdots\sigma_{j_p}+H\;.
\end{equation}
Hence --- in analogy to Ref.~\cite{tap} --- the TAP equations are
\begin{equation}
m_i=\tanh\beta\biggl({1\over{(p-1)!}}\sum_{j_2,\ldots,j_p}
J_{ij_2\cdots j_p}m'_{j_2}\cdots m'_{j_p}+H\biggr)\label{e4}\;,
\end{equation}
with $m'_j=m_j-\chi_{jj}\Delta'_j$, which is the magnetization of site
$j$ minus the Onsager--correction term (or cavity field). $\chi_{jj}$
is the susceptibility of spin $j$:
\begin{equation}
\chi_{jj}=\beta(1-m_j^2)\label{e5}
\end{equation}
and $\Delta h'_j$ is the field induced at site $j$ by the
magnetization of site $i$:
\begin{equation}
\Delta h'_j=m_i\cdot{1\over{(p-2)!}}\sum_{k_3,\ldots,k_p}
J_{ijk_3\cdots k_p}m_{k_3}\cdots m_{k_p}\;.\label{e6}
\end{equation}
Inserting this exression into eq.~(\ref{e4}) yields a rather
complicated structure for the TAP equations with terms up to order
$p$ in the couplings $J_{i_1,\cdots,i_p}$. In Appendix A we show
that neglecting terms of order $O(N^{-1})$ one gets
\begin{equation}
\tanh^{-1}(m_i)={\beta\over{(p-1)!}}\sum_{j_2,\ldots,j_p}
J_{ij_2\cdots j_p}m_{j_2}\cdots m_{j_p}+\beta H
-m_i{(\beta J)^2\over2}p(p-1)(1-q)q^{p-2}\;,
\label{e7}
\end{equation}
where we defined the self overlap of a paricular TAP solution
\begin{equation}
q={1\over N}\sum_i m_i^2\;.
\end{equation}
One can obtain eq.~(\ref{e7}) by differentiation of the free energy
functional $f$ with respect to $m_i$
\begin{eqnarray}
\beta f=&&{1\over 2N}\sum_i
\biggl\{ (1+m_i)\ln\biggl({{1+m_i}\over2}\biggr)+
(1-m_i)\ln\biggl({{1-m_i}\over2}\biggr)\biggr\}\nonumber\\
&&-{\beta\over N}\sum_{i_1<\cdots<i_p}J_{i_1\cdots i_p}m_{i_1}
\cdots m_{i_p} - {{(\beta J)^2}\over4}[(p-1)q^p-pq^{p-1}+1]\;.
\end{eqnarray}
The number of solutions $N_s$ of the TAP equations eq.~(\ref{e7})
is given by
\begin{equation}
N_s=N\int_0^1dq\,\int_{-1}^{+1}\prod_idm_i\,
\delta\biggl(Nq-\sum_i m_i^2\biggr)\,
\prod_i\delta(G_i)\,\vert{\rm det}\,{\bf A}\,\vert\;,
\end{equation}
where
\begin{eqnarray}
&G_i=&g(m_i)-{\beta\over{(p-1)!}}\sum_{j_2,\ldots,j_p}
J_{ij_2\cdots j_p}m_{j_2}\cdots m_{j_p}\;,\\
&g(m_i)=&\tanh^{-1}(m_i)+m_i{(\beta J)^2\over2}p(p-1)(1-q)q^{p-2}-
\beta H\;,\\
&{\bf A}_{ij}=&{{\partial G_i}\over{\partial m_j}}=a_i\delta_{ij}-
{\beta\over{(p-2)!}}\sum_{k_3,\ldots,k_p} J_{ijk_3\cdots k_p}
m_{k_3}\cdots m_{k_p}\;,\label{e13}\\
&a_i=&{1\over{1-m_i^2}}+{{(\beta J)^2}\over2}p(p-1)(1-q)q^{p-2}\;.
\end{eqnarray}
Following ref.~\cite{bray} we calculate $\langle N_s\rangle$ ---
which means the average of $N_S$ over the distribution of the
couplings (\ref{e2}) --- and discuss
the implication of the fact that one should rather introduce replicas
and perform the average $\ln \langle N_s\rangle$ in section
\ref{sec4}:
\begin{eqnarray}
\langle N_s\rangle=&N&
\int_{ }^{ } {{dq\,d\hat q}\over{2\pi}}
\int_{ }^{ } \prod_i {{dm_i\,d{\hat m}_i}\over{2\pi}}
\exp\biggl\{ \sum_i{\hat m}_i g(m_i)+
\hat q\biggl(Nq-\sum_i m_i^2\biggr)\biggr\}\\
&\cdot&\biggl\langle\exp\biggl\{{\beta\over{(p-1)!}}
\sum_{i_1,\ldots,i_p} J_{i_1\ldots j_p}{\hat m}_{i_1}m_{i_2}
\cdots m_{i_p}\biggr\}\cdot
{\rm det}({\bf A})\biggr\rangle\;.
\end{eqnarray}
The last factor can be written as
\begin{equation}
\langle\cdots\rangle=\prod_{i_1<\cdots<i_p}\exp
{{(\beta\tilde J)^2}\over{4(p-2)!}}\left(\sum_{\bf\pi}
{\hat m}_{{\bf\pi}(i_1)} m_{{\bf\pi}(i_2)} \cdots
m_{{\bf\pi}(i_p)}\right)^2\langle
{\rm det}\,{\bf A}'\rangle\;,\label{e15}
\end{equation}
where $\sum_{\bf\pi}$ is a sum over all permutations of $p$ different
integers $i_1,\ldots,i_p$ and
\begin{eqnarray}
{\bf A}'_{ij}=a_i&-&{\beta\over{(p-2)!}}\sum_{k_3,\ldots,k_p}
J_{ijk_3\cdots k_p}m_{k_3}\cdots m_{k_p}\nonumber\\
&+&{{(\beta J)^2 p}\over{(p-2)!N^{p-1}}} \sum_{k_3,\ldots,k_p}
\sum_{\bf\pi} {\hat m}_{{\bf\pi}(i)} m_{{\bf\pi}(j)}
m_{{\bf\pi}(k_3)}\cdots m_{{\bf\pi}(k_p)}\,m_{k_3}\cdots m_{k_p}\;.
\end{eqnarray}
The first sum yields a Gaussian random variable of order
$O(1/\sqrt{N})$, whereas the second sum produces three kinds
of terms: $N^{-1}({\hat m}_i m_j+m_i{\hat m}_j)q^{p-2}$,
$N^{-1}({\hat m}_i m_j+m_i{\hat m}_j)q^{p-4}M^2$ and
$N^{-1}m_i m_j q^{p-3}G$, where $M=N^{-1}\sum_i m_i$
and $G =N^{-1}\sum_i m_i{\hat m}_i$. They are all proportional to
$N^{-1}$ and can therefore be neglected, which implies
${\bf A}'={\bf A}$. The squared sum in eq.~(\ref{e15}) is
treated as follows
\begin{eqnarray}
&&\sum_{i_1,\ldots,i_p}\sum_{\bf\pi}{\hat m}_{{\bf\pi}(i_1)}
m_{{\bf\pi}(i_2)}\cdots
m_{{\bf\pi}(i_p)}\,{\hat m}_{i_1}\,m_{i_2}\cdots m_{i_p}\\
&=&(p-1)! \biggl\{N^{p-1}q^{p-1}\sum_i{\hat m}_i^2 +
(p-1) N^{p-2}q^{p-2}\biggl(\sum_i{\hat m}_i m_i\biggr)^2\biggr\}
\;.\nonumber
\end{eqnarray}
Using a Hubbard Stratonovi\c{c} transformation to linearize the
squared term we get
\begin{eqnarray}
\langle N_s\rangle=c&\cdot&\int dy\int dq d{\hat q}\int\prod_i
{{dm_id{\hat m}_i}\over2\pi}
\exp\biggl\{\sum_i{\hat m}_ig(m_i)
+{\hat q}\biggl(Nq-\sum_i m_i^2\biggr)\biggr\}\\
&&\exp\biggl\{{{(\beta J)^2}\over4}pq^{p-1}\sum_i{\hat m}_i^2
+y\sum_im_i{\hat m}_i-{{Ny^2}\over{(\beta J)^2p(p-1)q^{p-2}}}\biggr\}
\cdot\langle{\rm det}\,{\bf A}\rangle\;.\nonumber
\end{eqnarray}
In Appendix B we derive an expression for $\langle{\rm det}\,{\bf A}
\rangle$, the above integration will be done by steepest descent
(which becomes exact in the limit $N\rightarrow\infty$) and we are
left with
\begin{eqnarray}
\langle N_s\rangle=&&SP_{y,z,q,{\hat q}}\,\exp\,N\biggl({\hat q}q-
{{y^2}\over\lambda}+{{z^2}\over\lambda}\biggr)\\
&\cdot&\biggl\{\int_{-1}^{+1}dm\int_{-\infty}^{\infty}
{{d\hat m}\over{2\pi}}\exp\biggl[i\hat m g(m)
-\hat qm^2+{\mu\over2}q^{p-1}
(i\hat m)^2+ym\,i\hat m\biggr] (a-z)\biggr\}^N\;,\nonumber
\end{eqnarray}
where $SP$ means ``saddle point'' and
$\mu=(\beta J)^2p/2$, $\lambda=2\mu(p-1)q^{p-2}$ and $a=(1-m^2)^{-1}+
(1-q)\lambda/2$. For simplicity we set the external field $H$ to zero
from now on. Introducing the variables used in ref.~\cite{bray}
\begin{eqnarray}
B&=&z-\lambda(1-q)/2\;,\nonumber\\
\Delta&=&y+\lambda(1-q)/2
\end{eqnarray}
and performing the integration over $\hat m$ (which is Gaussian)
we get
\begin{eqnarray}
\langle N_s\rangle&=&SP_{B,\Delta,q,\hat q}\exp\,N
\biggl[-q\hat q-(B+\Delta)
(1-q)+{{B^2-\Delta^2}\over{2\mu(p-1)q^{p-2}}}+\ln I\biggr]\;,\\
I&=&\int_{-1}^{+1}{dm\over\sqrt{2\pi\mu q^{p-1}}}\biggl(
{1\over{1-m^2}}+B\biggr)\exp\biggr\{-{{(\tanh^{-1}(m)-\Delta m)^2}
\over{2\mu q^{p-1}}}+\hat q m^2\biggr\}\;,\label{e22}
\end{eqnarray}
which reduces in the special case $p=2$ to eq.~(15--17) of
ref.~\cite{bray}.
The saddle point equations again admit the solution $B=0$,
which we adopt here. We define $\sigma=\sqrt{\mu q^{p-1}}$,
$\tilde\Delta = \Delta/\sigma$ and use the variable substitution
$x=\tanh^{-1}(m)/\sigma$ to proceed further. Then the saddle point
equations for $q$, $\tilde\Delta$ and $\hat q$ read
\begin{eqnarray}
q&=&{1\over I'}\int_0^\infty dx\,[\tanh(\sigma x)]^2\,
\exp[L(\sigma,\tilde\Delta,
\hat q)]\;,\nonumber\\
{\tilde\Delta}&=&{{p-1}\over{pq}}
{1\over I'}\int_0^\infty dx\,x\,\tanh(\sigma x)\,
\exp[L(\sigma,\tilde\Delta,
\hat q)]\,-\,p^{-1}\mu(1-q)q^{p-2}(p-1)\;,\label{e23}\\
{\hat q}&=&\biggl\{
{1\over I'}\int_0^\infty dx\,[x-\tilde\Delta\tanh(\sigma x)]^2\,
\exp[L(\sigma,\tilde\Delta,
\hat q)]\,-\,1\biggr\}\,{{p-1}\over{2q}}+\sigma\tilde\Delta
+{{p-2}\over{p-1}}{{\tilde\Delta}^2\over2}\;,\nonumber
\end{eqnarray}
where
\begin{eqnarray}
L(\sigma,\tilde\Delta,\hat q)&=&-{1\over2}
[x-\tilde\Delta\tanh(\sigma x)]^2
+\hat q\,[\tanh(\sigma x)]^2\;,\nonumber\\
I'&=&\int_0^\infty dx\,\exp\,[L(\sigma,\tilde\Delta,\hat q)]\;.
\end{eqnarray}
Since $\tilde\Delta$ is well--behaved for $T\rightarrow0$ the quantity
$\Delta$ diverges, which is the main reason for this substitution. For
the numerical solution of the saddle point equations it is also
advisable to use $(\hat q-\Delta)$ instead of $\hat q$, since the
latter also diverges for $T\rightarrow0$.

\section{Results}\label{sec3}

First we look at the the zero temperature
limit. For $\beta\rightarrow\infty$ the quantity $\mu$ will diverge
also and therefore $\sigma\rightarrow\infty$, too. Hence we may
replace $\tanh(\sigma x)$ by $1$ within the integrals, since they
only extend over $x>0$. This yields (as expected) $q=1$ and
furthermore, after a shift $y=x-\tilde\Delta$
\begin{equation}
I'=e^{\hat q}\,\int_{-\tilde\Delta}^\infty dy\,e^{-y^2/2}
\end{equation}
and hence
\begin{equation}
\tilde\Delta = (p-1)\cdot{{e^{-{\tilde\Delta}^2}}\over{
\int_{-\tilde\Delta}^\infty dy\,e^{-y^2/2}}}\;.\label{e26}
\end{equation}
With the above defined quantities (and $I=I'\sqrt{2/\pi}$) we get
for $T=0$
\begin{equation}
\langle N_s\rangle
=\exp\,N\left\{\ln 2-{{\tilde\Delta}^2\over{2(p-1)}}
+\ln\int_{-\tilde\Delta}^\infty dy\,e^{-y^2/2}\right\}\;,
\end{equation}
where $\tilde\Delta$ has to be determined via eq.~(\ref{e26}). This
is in complete agreement with what was found by Gross and M\'ezard
ref.~\cite{gross}.
For $T=0$ the number of TAP--solutions increases monotonically from
$\ln\,N_s=0.1992\,N$ for $p=2$ (see also \cite{tanaka})
to $\langle N_s\rangle\approx2^N$ for $p\rightarrow\infty$
(the random energy model).

The exploration of the behavior of $N_s$ for nonvanishing temperatures
has to be done numerically. The success of numerical methods for
solving nonlinear equations like eq.~(\ref{e23}) rely heavily on the
quality of an initial--guess. Therefore we started at $T=1-t$ with
$t\ll1$ (we set $J=1$ from now on) for $p=2$, where one has an
analytic expression for $q$, $\tilde\Delta$ and $\hat q$ in form
of a power series in $t$ \cite{bray}.
Then we decreased the temperature by small steps $\delta T$, always
using $q(T)$, $\Delta(T)$ and $\hat q (T)$ as an initial guess for
$q(T-\delta T)$, $\Delta(T-\delta T)$ and $\hat q (T-\delta T)$.
This was done down to very small temperatures and then we fixed the
temperature and increased $p$ by small steps $\delta p$ up to $p=3$
using the same procedure. Finally we fixed $p=3$ and increased the
temperature again, up to a value, where the solution disappeared
discontinuously. In the same way we solved eq.~(\ref{e23}) for
higher values of $p$. Numerically it is much more difficult
to start with $p=2$ and a temperature near
$T=1$ and then to increase $p$ slowly, since $q=\tilde\Delta=
\hat q=0$ is always a stable solution for any temperature as long
as $p>2$. By the method we used we were sure to be on the same
solution--branch as the low--temperature solution for $p=2$. We
did not find other non--trivial solutions of eq.~(\ref{e23})
although they might exist.

The result for $\alpha(T)=N^{-1}\ln\langle N_s\rangle$ is shown in
fig.~\ref{fig1}. At $T_c(p)$ the factor $\alpha(T)$ jumps
discontinuously for any $p>2$. Furthermore one observes that as
long as there is an exponentially large number of TAP--solutions
it increases monotonically with $p$ for a fixed temperature.
With increasing $p$ the critical temperature $T_c(p)$ first decays
from $T_c=1$ for the SK model ($p=2$) to a value slightly
below the critical temperature $T_c^{\rm RE}$ of the random energy
model ($p\to\infty$: $T_c^{\rm RE}=(2\sqrt{\ln\,2})^{-1}\approx0.60$
\cite{derrida}).
For $p>4$ it increases slowly to that value. Also
$\langle N_s\rangle$ approaches for $p\to\infty$ the form predicted
by the random energy model, which is a step function:
$\alpha(T)=\ln\,2\cdot\theta(T_c^{\rm RE}
-T)$. For $p\to\infty$ nearly every spin--configuration of the system
is a solution of the TAP--equations \cite{gross}.

In fig.~\ref{fig2} we plotted the result of the mean--squared
magnetization
$q=\langle m^2\rangle$ of the TAP--solutions in dependence of the
temperature. The curve for the SK model ($p=2$) resembles that of
the EA order parameter $q_{\rm EA}$ (or $q(x=1)$ in Parisi's theory
\cite{parisi}) --- but we point out that they are not the same, since
we performed a white average over the solutions (see section
\ref{sec1}). As long as there is an exponentially
large number of TAP--solutions $q$ increases monotonically with $p$
for a fixed temperature. Again we observe that in the limit
$p\to\infty$ the result for the random energy model, where
$q=1$ below $T_c^{\rm RE}$ is approached. This means that in nearly
every configuration of the system all spins are frozen below $T_c$.

{}From the last figure one recognizes that solutions of the
TAP--equations describe configurations of the system that are
frozen to a large
extent already for rather small values of p. This means that the
spins are more or less fixed to values $+1$ or $-1$ and do not fluctuate
significantly below $T_c(p)$. This feature is expected in the
limit $p\to\infty$, but it is rather surprising that it is a good
approximation for $p$ as small as 5.
In \cite{gross} it was argued that an expansion of the free energy
and the order parameter function $q(x)$ around $p=\infty$ might be
rapidly convergent, which means that already for rather small values
of $p>2$ the features of the random energy model might dominate
the physics of p--spin interaction spin glasses. Our results support
this conjecture.

\section{Discussion}\label{sec4}

After reducing the problem of calculating the number $N_s(T)$ of
TAP solutions for p--spin interaction spin glasses to a set of
self--consistency equations, we showed that below a critical
temperature this number becomes exponentially large: $N_s(T)\propto
\exp[\alpha(T)\,N]$. For $p>2$ the factor $\alpha(T)$ jumps
discontinuously at the critical temperature $T_c$. This is
consistent with existing mean--field theories of p--spin interaction
spin glasses, where the discontinuity is observed in the EA--order
parameter $q_{\rm EA}$. For increasing $p$ the critical temperature
approaches the (exactly known) value for the random energy
model very rapidly and the shape of $\alpha(T)$ approaches
the step function of the random energy model.

We also observe that the mean squared magnetization
$\langle m^2\rangle$ of the TAP--solutions behaves similarly to
$q_{\rm EA}$ and becomes equal to it in the limit $p\to\infty$.
Nevertheless it should be noted that
for finite $p$, $\langle m^2\rangle\ne q_{\rm EA}$, since we
performed a white average, whereas $q_{\rm EA}$ can only be obtained
by attributing a Boltzman weight to the TAP--solutions. Furthermore,
the discontinuity in $\alpha(T)$ is reminiscent of the jump in a
quantity called ``complexity'' or ``configurational entropy'' in the
context of metastable states in p--state Potts glasses with $p>4$
\cite{trkwol}.

The latter work, combined with the observation that the mean--field
theory for p--spin interaction spin glasses with $p>2$ and that
for p--state Potts glasses with $p>4$ seem to be in the same
universality class \cite{trk},
gives us reason to believe that performing the average
over $N_s$ --- as we did --- instead over $\ln(N_s)$ yields the
right factor $\alpha(T)$ at least in a small temperature regime
around $T_c$. In the Potts--case it was shown \cite{trkwol} that
in a temperature
regime $T_K<T<T_g$ ($T_g$ is the temperature, where the above
mentioned configurational entropy jumps discontinuously), the
TAP--solutions only have self--overlap.
This means that calculating $\ln\langle N_s\rangle$ via replicas
requires only diagonal order--parameters (in replica space) for
$T>T_K$ \cite{trkwol}.
Hence the behavior of $\alpha(T)$ around $T_c$ --- especially the
discontinuity --- will not be changed if one performs the correct
average over $\ln\langle N_s\rangle$. At a lower temperature
$T_K<T_c$, where the true equilibrium phase transition takes place,
the picture might change slightly. It is desirable to undertake
more detailed investigations on the above mentioned points and
work on the TAP--approach to the thermodynamics of p-spin
interaction spin glasses is in progress \cite{hr2}.

\section{Acknowledgements}
The author is grateful to T.\ R.\ Kirkpatrick for numerous
valuable discussions and would like to thank the ITP and
the Physics Department of the University of California at Santa Cruz,
where the last part of this work was done, for their kind hospitality.
Furthermore he acknowledges financial support from the Deutsche
Forschungsgemeinschaft (DFG).

\appendix{ }
Here we derive eq.~(\ref{e7}) of the main text. Remembering
eq.~(\ref{e4}--\ref{e6}) we have
\begin{equation}
{\beta\over{(p-1)!}} \sum_{j_2,\ldots,j_p}J_{ij_2\cdots j_p}
m_{j_2}'\cdots m_{j_p}'\,=\,A-B+C\;,\label{a1}
\end{equation}
where we define
\begin{eqnarray}
A&=&{\beta\over{(p-1)!}}\sum_{j_2,\ldots,j_p}J_{ij_2\cdots j_p}
m_{j_2}\cdots m_{j_p}\;,\\
B&=&{\beta^2\over{(p-2)!^2}}\sum_{j_2,\ldots,j_p}J_{ij_2,\ldots,j_p}
m_i(1-m_{j_2}^2)\,m_{j_2}\cdots m_{j_p}\cdot
\sum_{k_3,\ldots,k_p}J_{ij_2k_3\cdots k_p}m_{k_3}\cdots m_{k_p}\;.
\end{eqnarray}
The remaining part $C$ contains terms of higher order than second
within the couplings. Note that we have already used the permutation
symmetry of the couplings for the expression $B$. It can be written as
\begin{equation}
B={{\beta^2 m_i}\over(p-2)!^2}\sum_j(1-m_j^2)
\biggl(\sum_{k_3,\ldots,k_p}J_{ijk_3\cdots k_p}m_{k_3}\cdots m_{k_p}
\biggr)^2\;.
\end{equation}
The sum, which is squared is a sum over $M=N^{p-2}$ independent random
variables $X_{j,1},\ldots,X_{j,M}$, whose variance is
$\langle X^2\rangle\propto1/N^{p-1}$. It can be splitted into two
parts:
\begin{equation}
S_j=\left(\sum_{n=1}^M X_{j,n}\right)^2=
\underbrace{\sum_n X_{j,n}^2}_{\textstyle=:{\overline{S}}_j}+
\underbrace{\sum_{n\ne m} X_{j,n}X_{j,m}}_{\textstyle=:\delta S_j}\;.
\end{equation}
The term $\overline{S_j}$ is a sum over positive random numbers
of order $1/N^{p-1}$ and therefore yields a quantity of order
$O(M/N^{p-1})=O(1/N)$, whereas $\delta S_j$ yields (by using the
central limit theorem) a Gaussian random variable with mean zero
and variance $M^2\langle X^2\rangle^2=O(1/N^2)$. Hence $B$ is
\begin{equation}
B={{\beta^2 m_i}\over(p-2)!^2}\sum_j(1-m_j^2)
[\overline{S}_j+\delta S_j]=\overline{B}+\delta B\;.
\end{equation}
$\overline{B}$ is a term of order $O(N\cdot\overline{S})=O(1)$,
whereas $\delta B$ is again a Gaussian variable with zero mean
and variance $N\cdot\langle\delta S_j^2\rangle\propto1/N$, which
can be neglected with respect to $\overline{B}$. This leaves us
with
\begin{equation}
B={{\beta^2 m_i}\over(p-2)!^2}\sum_j(1-m_j^2)
(p-2)!\sum_{k_3,\ldots,k_p}J_{ijk_3\cdots k_p}^2m_{k_3}^2
\cdots m_{k_p}^2\;,
\end{equation}
the factor $(p-2)!$ stems from the permutation symmetry of the
couplings. Using
\begin{equation}
N^{-1}\sum_{k_p}J_{ijK_3\ldots k_p}^2 m_{k_p}^2\approx\langle
J_{ijk_3\cdots k_p}^2\rangle\langle m_{k_p}^2\rangle=
{{J^2 p!}\over{2N^{p-1}}}q\;,
\end{equation}
where we neglected again fluctuations around the average, which are
of lower order in $N$, we finally end up with
\begin{equation}
B={{(\beta J)^2}\over2}p(p-1)(1-q)q^{p-2}\;.
\end{equation}
The part $C$ on the r.h.s.\ of eq.~(\ref{a1}) is a sum over terms
($s\ge2$)
\begin{equation}
T=\sum_{j_1,\ldots,j_s}(1-m_{j_1}^2)\cdots(1-m_{j_s}^2)
\left(\sum_{k_{s+1},\ldots,k_p}J_{ij_1\cdots j_sk_{s+1}\cdots k_p}
m_{k_{s+1}}\cdots m_{k_p}\right)^{s+1}\;.
\end{equation}
The order of $T$ is given by
\begin{equation}
O(T)=O\biggl(N^s\biggl[\sum_{n=1}^{M_s} x_n\biggr]^{s+1}\biggr)\;,
\end{equation}
where $M_s=N^{p-(s+1)}$ and $X_n$ are again independent random
variables with zero mean and variance $\langle X^2\rangle=1/N^{p-1}$.
Therefore the order of $T$ cannot be greater than $O(N^{-s^2})$,
which proves the correctness of eq.~(\ref{e7}) up to order $O(1/N)$.

\appendix{ }
Here we calculate $\langle{\rm det}\,{\bf A}\rangle$ by using the
identity
\begin{equation}
{\rm det}\,{\bf A}=\lim_{n\rightarrow-2}\int\prod_{i=1}^N
\prod_{\alpha=1}^n{{d\xi_i^\alpha}\over\sqrt{2\pi}}
\exp\biggl\{-{1\over2}\sum_{i\alpha}\xi_{i\alpha}{\bf A}_{ij}
\xi_{j\alpha}\biggr\}\;.
\end{equation}
{}From eq.~(\ref{e13}) in the main text we have then
\begin{eqnarray}
\langle&&{\rm det}\,{\bf A}\rangle=
\biggl\langle\int\prod_{i,\alpha} {{d\xi_i^\alpha}\over\sqrt{2\pi}}
\exp\biggl\{-{1\over2}\sum_{i,\alpha}a_i\xi_{i\alpha}^2
+{\beta\over{2(p-2)!}}\sum_\alpha\sum_{i_1,\ldots,i_p}
J_{i_1\cdots i_p}\xi_{i_1\alpha}\xi_{i_2\alpha}m_{i_3}\cdots
m_{i_p}\biggr\}\biggr\rangle\nonumber\\
 & & \\
&&=
\int\prod_{i,\alpha} {{d\xi_i^\alpha}\over\sqrt{2\pi}}
\exp\biggl\{-{1\over2}\sum_{i,\alpha}a_i\xi_{i\alpha}^2
+{(\beta\tilde J)^2\over{16(p-2)!^2}}\sum_{i_1<\cdots<i_p}
\biggl(\sum_{{\bf\pi},\alpha}\xi_{{\bf\pi}(i_1)\alpha}
\xi_{{\bf\pi}(i_2)\alpha}m_{{\bf\pi}(i_3)}\cdots
m_{{\bf\pi}(i_p)}\biggr)^2\biggr\}\;,\nonumber
\end{eqnarray}
where $\sum_{\bf\pi}$ means again the sum over all permutations of
$p$ different integers $i_1,\ldots,i_p$. After some algebra we get
\begin{eqnarray}
\langle{\rm det}\,{\bf A}\rangle=
\int&&\prod_{i,\alpha}{{d\xi_i^\alpha}\over\sqrt{2\pi}}
\exp\bigg\{-{1\over2}\sum_{i,\alpha}a_i\xi_{i\alpha}^2\nonumber\\
&&+N {(\beta J)^2\over16}p(p-1)
\sum_{\alpha,\gamma}\biggl[2q^{p-2}\biggl({1\over N}
\sum_i\xi_{i\alpha}\xi_{i\gamma}\biggr)^2\label{astar}\\
&&\quad+4(p-2)q^{p-3}\biggl({1\over N}\sum_i\xi_{i\alpha}\xi_{i\gamma}
\biggr)\biggl({1\over N}\sum_i\xi_{i\alpha}m_i\biggr)
\biggl({1\over N}\sum_i\xi_{i\gamma}m_i\biggr)\nonumber\\
&&\quad+(p-2)(p-3)q^{p-4}
\biggl({1\over N}\sum_i\xi_{i\alpha}m_i\biggr)^2
\biggl({1\over N}\sum_i\xi_{i\gamma}m_i\biggr)^2\biggr]\bigg\}\;.
\nonumber
\end{eqnarray}
To see that one can neglect the last two terms one introduces
$\delta$--funtions for
\begin{eqnarray}
z_{\alpha\gamma}&=&{1\over N}\sum_i\xi_{i\alpha}\xi_{i\gamma}\;,
\nonumber\\
x_\alpha&=&{1\over N}\sum_i\xi_{i\alpha}m_i
\end{eqnarray}
with the help of conjugate variables ${\hat z}_{\alpha\gamma}$ and
${\hat x}_\alpha$ respectively (we do not pay attention to the fact
that $z_{\alpha\gamma}=z_{\gamma\alpha}$ because it does not matter
for this discussion).
\begin{eqnarray}
\langle{\rm det}\,{\bf A}\rangle=
N^2&&\int\prod_{i,\alpha}{{d\xi_i^\alpha}\over\sqrt{2\pi}}
\int\prod_{\alpha,\gamma}{{dz_{\alpha\gamma}d{\hat z}_{\alpha\gamma}}
\over{2\pi}}\int\prod_\alpha{{dx_\alpha d{\hat x}_\alpha}\over{2\pi}}
\exp\biggl\{-{1\over2}\sum_{i,j}\sum_{\alpha,\gamma}\xi_{i\alpha}
M_{ij,\alpha\gamma}\xi_{j\gamma}\biggr\}\nonumber\\
&&\exp\,N\biggl\{\sum_{\alpha,\gamma}
z_{\alpha\gamma}{\hat z}_{\alpha\gamma}
+\sum_\alpha{\hat x}_\alpha x_\alpha+{{(\beta J)^2}\over2}
p(p-1)q^{p-2}\sum_{\alpha\gamma} z_{\alpha\gamma}^2\biggr\}\;,
\end{eqnarray}
where
\begin{eqnarray}
M_{ij,\alpha\gamma}=&&([a_i-{\hat z}_{\alpha\alpha}]
\delta_{\alpha\gamma}-{\hat z}_{\alpha\gamma})
\delta_{ij}\nonumber\\
&&-{{(\beta J)^2}\over8}p(p-1)(p-2)\{4q^{p-2}z_{\alpha\gamma}
+(p-3)x_\alpha x_\gamma q^{p-4}\}\cdot{1\over N}m_i m_j\;.
\end{eqnarray}
The last term is of order $O(1/N)$ and corresponds to the two terms
under discussion. Therefore they will be dropped and we get from
eq.~(\ref{astar}) after performing a Hubbard--Stratonovic
transformation
\begin{equation}
\langle{\rm det}\,{\bf A}\rangle=
\int\prod_{\alpha\gamma}
{{dz_{\alpha\gamma}}\over\sqrt{2\pi\lambda N^{-1}}}
\int\prod_{i,\alpha}{{d\xi_i^\alpha}\over\sqrt{2\pi}}
\exp\biggl\{-{1\over2}\sum_{i,\alpha}a_i\xi_{i\alpha}^2-N
\sum_{\alpha,\gamma}{{z_{\alpha\gamma}^2}\over{2\lambda}}+{1\over2}
\sum_{\alpha,\gamma}\sum_i z_{\alpha\gamma}\xi_{i\alpha}
\xi_{i\gamma}\biggr\}\;,
\end{equation}
where $\lambda=(\beta J)^2p(p-1)q^{p-2}$. We calculate the integral
over $z_{\alpha\gamma}$ by steepest descent and choose the diagonal
saddle point $z_{\alpha\gamma}=z\delta_{\alpha\gamma}$. We calculated
$\langle{\rm det}\,{\bf A}\rangle$ also with the help of
Grassman--variables
instead of replicas (see ref.~\cite{domyoung}) and checked in this way
the correctness of this saddle--point. Then we obtain
\begin{eqnarray}
\langle{\rm det}\,{\bf A}\rangle&=&c\cdot
\int\prod_{i,\alpha}{{d\xi_i^\alpha}\over\sqrt{2\pi}}
\exp\biggl\{-{1\over2}\sum_{i,\alpha}(a_i-2z)\xi_{i\alpha}^2-N
\sum_\alpha{{z^2}\over{2\lambda}}\biggr\}\nonumber\\
&=&c\cdot\prod_i(a_i-2z)^{-n/2}\exp\{-{{Nnz^2}\over2\lambda}\}
\nonumber\\
&\stackrel{n\rightarrow-2}{\longrightarrow}&
c\cdot\prod_i(a_i-2z)\exp\biggl\{{{Nz^2}\over\lambda}\biggr\}\;,
\end{eqnarray}
where $c$ is a prefactor and $z$ has to be determined variationally.

\begin{references}

\bibitem[*]{hrieger} Present address: Physics Department,
University of California, Santa Cruz, CA 95064.

\bibitem{by} K.~Binder, and A.~P.~Young, Rev.\ Mod.\ Phys.\ {\bf 58},
801 (1986).

\bibitem{tap} D.~J.~Thouless, P.~W.~Anderson, and R.~G.~Palmer,
Phil.\ Mag.\ {bf 35}, 593 (1977).

\bibitem{sk} D.~Sherrington, and S.~Kirkpatrick, Phys.\ Rev.\ Lett.\
{\bf 35}, 1972 (1975).

\bibitem{bray} A.\ J.\ Bray, and M.\ A.\ Moore, J.\ Phys.\ C {\bf 13},
L469 (1980).

\bibitem{parisi} G.~Parisi, J.\ Phys.\ A {\bf 13}, 1101 (1980);
{\bf 13}, 1887 (1980); {\bf 13}, L115 (1980).

\bibitem{domyoung} C.~de Dominicis, and A.~P.~Young, J.\ Phys.\ A
{\bf 16}, 2063 (1983).

\bibitem{derrida} B.~Derrida, Phys.\ Rev.\ Lett.\ {\bf 45}, 79 (1980);
Phys.\ Rev.\ B {\bf 24}, 2613 (1981).

\bibitem{gross} D.~J.~Gross, and J.~M\'ezard, Nucl.\ Phys.\ B
{\bf 240}, 431 (1984).

\bibitem{gardner} E.~Gardner, Nucl.\ Phys.\ B {\bf 257}{$\;[FS14]$},
747 (1985).

\bibitem{sommers} A.~Crisanti, and H.~J.~Sommers, Z.\ Phys. B,
{\bf 87}, 341 (1992).

\bibitem{hrtrk} H.~Rieger and  T.~R.~Kirkpatrick, Phys.\ Rev.\ B
{\bf 45}, 9772 (1992).

\bibitem{hr} H.~Rieger, Physica A {\bf 184}, 279 (1992).

\bibitem{trk} T.~R.~Kirkpatrick, and D.~Thirumalai, Phys.\ Rev.\ B
{\bf 36}, 5388 (1987).

\bibitem{trkwol} T.~R.~K519 2015 li 1521